\documentclass[12pt]{article}

\usepackage{newtxtext,newtxmath}

\usepackage{graphicx}

\usepackage[letterpaper,margin=1in]{geometry}

\renewenvironment{abstract}
	{\quotation}
	{\endquotation}
	
\date{}

\makeatletter
\renewcommand{\fnum@figure}{\textbf{Figure \thefigure}}
\renewcommand{\fnum@table}{\textbf{Table \thetable}}
\makeatother

\usepackage{scicite}

\usepackage{url}

\def\scititle{Unveiling the nature of the knee in the cosmic ray energy spectrum}
\title{\bfseries \boldmath \scititle}

\author{Huihai~He$^{1,2,3}$,
	   Hengying~Zhang$^{4}$,
	   Qinyi~Cheng$^{1,2}$,
	   Lingling~Ma$^{1,3\ast}$,
	   Cunfeng~Feng$^{5}$\and
	   \scriptsize$^{1}$Key Laboratory of Particle Astrophysics, Institute of High Energy Physics, Chinese Academy of Sciences, Beijing \& 100049, China.\and
	   \scriptsize$^{2}$University of Chinese Academy of Sciences, Beijing \& 100049, China.\and
	   \scriptsize$^{3}$Tianfu Cosmic Ray Research Center, Chengdu \& 610000, China.\and
	   \scriptsize$^{4}$School of Physics and Astronomy, Yunnan University, Kunming \& 650091, China.\and
	   \scriptsize$^{5}$Institute of Frontier and Interdisciplinary Science, Shandong University, Qingdao \& 266237, China.\and
	   \scriptsize$^\ast$Corresponding author. Email: llma@ihep.ac.cn}

\begin{document}

\maketitle

\begin{abstract} \bfseries \boldmath
The knee-like structure around 4~PeV is the most striking feature in the cosmic ray energy spectrum, whose origin remains enigmatic.
We propose a novel concept of the total logarithmic mass energy spectrum to characterize the knee, taking into account LHAASO measurements of the all-particle energy spectrum and the mean logarithmic mass.
The predominant role of proton in the knee formation is unearthed.
The case of a mass-dependent knee is ruled out with a significance of 22.9$\sigma$ and the rigidity-dependent knee feature is revealed.
An ankle-like structure stemming from the excess of iron is discovered at $9.7 \pm 0.2$~PeV with a significance of 25.9$\sigma$.
Our findings pierce the mist of the puzzling knee for the first time since its discovery.
\end{abstract}

\noindent
Cosmic rays (CRs) are highly energetic particles arriving from outer space. 
The energy spectrum and chemical composition of CRs can furnish insights into their astrophysical origin, acceleration mechanism and propagation regime \cite{Amato_2018}.
The energy spectrum of CRs extends an immense range from 10$^{9}$~eV to 10$^{20}$~eV, exhibiting a steeply falling near-power-law distribution of ${\mathrm{d} N} / {\mathrm{d} E} \propto E^{\gamma}$.
Several features described as changes of the differential spectral index $\gamma$ have been observed in the spectrum.
The most prominent feature christened the knee refers to a softening of the spectrum at approximately 4~PeV \cite{Kulikov_1959}, with the spectral indices of -2.7 and -3.1 before and beyond the knee.
This knee-like structure has been supported by various experiments, such as AS$\gamma$ \cite{Amenomori_2008}, KASCADE \cite{Antoni_2005}, ARGO-YBJ \cite{Montini_2016}, CASA-MIA \cite{Glasmacher_1999} and IceCube/IceTop \cite{Aartsen_2019,Aartsen_2020}.
Despite efforts spanning over 60 years since its discovery, the origin of the knee remains shrouded in mystery.

Over the past few decades, numerous models have been proposed to shed light on the properties of the CR knee \cite{Kachelriess_2019}.
In general, those explanations can be divided into three categories.
The first class of models attributes the knee to acceleration or propagation processes of Galactic CRs \cite{Erlykin_2001}, consequently leading to the so-called rigidity-dependent (also $Z$-dependent) knee.
In this scenario, the cutoff energies of different elements are naturally proportional to the atomic number $Z$ \cite{Peters_1961}.
According to the second class of models, new physics is responsible for the CR knee \cite{Nikolsky_1995}, which triggers the so-called mass-dependent (also $A$-dependent) knee.
In this case, a cutoff energy for each element proportional to its mass $A$ is assumed \cite{Horandel_2004}.
The last class is neither $Z$-dependent nor $A$-dependent, thus allowing fluxes of all the CR nuclei to decrease at the same break point in the knee region.
Unquestionably, investigating the energy spectra of individual species is crucial for distinguishing between the three classes of models and bringing to light the nature of the CR knee.

\subsection*{Total logarithmic mass energy spectrum}
Owing to sharp decrease in the flux of primary particles, measurements of CRs around the knee region can only be carried out through observations of extensive air showers (EASs) initiated by CRs interacting with atmospheric nuclei \cite{Auger_1939}.
Although such ground-based indirect experiments boast large effective detection areas \cite{Bose_2022}, information about the primary energy and chemical composition of incident particles gets obscured in cascade development, necessitating the assumption of composition models for the energy reconstruction \cite{Haungs_2003}.
Additionally, due to limited knowledge about high-energy hadronic interaction mechanisms, models extrapolated from lower-energy accelerator data are also necessarily involved in numerical simulations of EASs \cite{Fedynitch_2012}.
The theoretical uncertainties in these models inevitably bring about large systematic uncertainties of measurement results \cite{Parsons_2011}.
Furthermore, shower fluctuations often complicate the discrimination between different primaries \cite{Bortolato_2023}.
When it comes to measuring the proton spectrum, the selection efficiency of proton-induced events is sacrificed to mitigate the contamination from primaries heavier than proton. 
Given the contamination from both proton and heavier nuclei, measuring the helium spectrum is even more challenging. 
So the selection of proton plus helium \cite{Bartoli_2015}, instead of individual proton and helium contributions, offers an alternative method to explore the helium spectrum. 
However, subtracting the proton contribution from the energy spectrum of light elements is faced with the implication of the proton-to-helium ratio.
As a result, none of ground-based experiments have precisely measured the spectra of individual elements in the knee region so far.

Aimed at illuminating the intrinsic characteristics of the CR knee, a brand-new method is presented here, ushering in the novel concept of the total logarithmic mass energy spectrum.
In previous works, the mean logarithmic mass $\langle \ln A \rangle$ (denoted by $m$ thereinafter) is a commonly used quantity to characterize the CR composition, which is defined as 
\begin{equation}
           m = \frac{\sum_i f_i \ln A_i}{\sum_i f_i},
          \label{eq:lnA}
\end{equation}
where $A_i$ denotes the mass number of the CR nuclei, $f_i$ denotes the flux of different elements from proton to iron, and $F = \sum_i f_i$ denotes the all-particle energy spectrum flux of CRs. 
In this work, the total logarithmic mass energy spectrum flux $M$ is defined as 
\begin{equation}
           M = \sum_i f_i \ln A_i.
          \label{eq:tlm}
\end{equation}
From another perspective, $M$ can be directly calculated as
\begin{equation}
           M = F \times m.
          \label{eq:TLM}
\end{equation}
In essence, the total logarithmic mass energy spectrum can be perceived as a weighted spectrum, where contributions of different elements are scaled by their respective logarithmic masses.
Notably, the logarithmic mass of proton is 0, meaning that proton serves as a non-contributory ingredient for the total logarithmic mass energy spectrum.
Indeed, if proton dominates the knee, the total logarithmic mass energy spectrum will exhibit no distinct structure around the knee energy; if other nuclei assert dominance over the knee, the total logarithmic mass energy spectrum will display a knee-like structure similar to that observed in the all-particle energy spectrum.
Therefore, the total logarithmic mass energy spectrum provides a key to unlocking the puzzle of whether proton has dominion over the CR knee.
Meanwhile, heavier nuclei contribute more significantly to the total logarithmic mass energy spectrum with greater weights.
In particular, iron has a weight factor of $\ln 56$, which is the largest among all the CR elements.
To some extent, the total logarithmic mass energy spectrum is more sensitive to the flux contributions from heavy elements.
Thus, delving into the total logarithmic mass energy spectrum before or beyond the knee can also facilitate access to the spectral structures of heavy nuclei.

\begin{figure}
          \centering
          \includegraphics[width=0.6\textwidth]{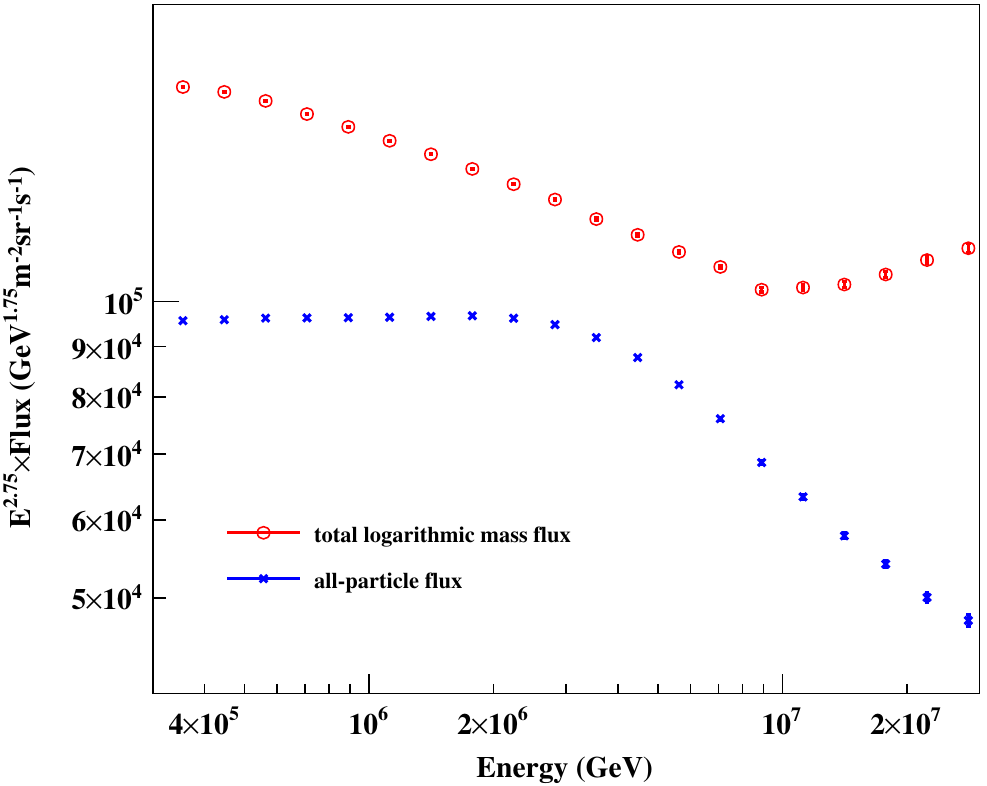}
          \caption{\textbf{The total logarithmic mass energy spectrum and the all-particle energy spectrum from 0.3 to 30~PeV.} 
                        The red hollow circles represent the total logarithmic mass energy spectrum flux $M$ multiplied by E$^{2.75}$, which is calculated with LHAASO measurement results \cite{Zhang_2024}.
                        The blue crosses represent the all-particle energy spectrum flux $F$ multiplied by E$^{2.75}$, which is also obtained from LHAASO measurement results \cite{Zhang_2024}.}
          \label{fig:tlm}
\end{figure}

Benefiting from its high altitude and simultaneous measurements of both electromagnetic particles and muons, the Large High Altitude Air Shower Observatory (LHAASO) is able to perform calorimetric measurements, thereby greatly reducing the dependence of energy reconstruction on composition models and hadronic interaction models \cite{Zhang_2022}. 
Thanks to the significant advantages of LHAASO, the all-particle energy spectrum and the mean logarithmic mass of CRs around the knee energy have been measured with unprecedented accuracy \cite{Zhang_2024}.
Compared to other experiments like AS$\gamma$ \cite{Amenomori_2008}, KASCADE \cite{Antoni_2005}, ARGO-YBJ \cite{Montini_2016}, CASA-MIA \cite{Glasmacher_1999} and IceCube/IceTop \cite{Aartsen_2019,Aartsen_2020}, LHAASO distinguishes itself by achieving remarkably low systematic uncertainties.
Hence, the published LHAASO results are utilized to obtain the total logarithmic mass energy spectrum from 0.3 to 30~PeV (Fig.~\ref{fig:tlm}).
Obviously, the total logarithmic mass energy spectrum exhibits a gradual decrease before 10~PeV and demonstrates an upward bend after 10~PeV.
An ankle-like structure is discovered at $9.7 \pm 0.2$~PeV and iron is responsible for it (see the penultimate section).

\subsection*{Proton contribution to the knee}
If all the CR elements manifested themselves in the same mechanism before and beyond the knee, the descending trends of the all-particle energy spectrum, the mean logarithmic mass and the total logarithmic mass energy spectrum would remain unchanged in the knee region. 
In this respect, the expected flux $M''$ of the total logarithmic mass energy spectrum can be expressed as \cite{methods}
\begin{equation}
           M'' = M' - \Delta F \times m_{\Delta F},
          \label{eq:expM}
\end{equation}
where $M'$ is the extrapolated flux of the total logarithmic mass energy spectrum, $\Delta F$ is the flux variation in the all-particle energy spectrum and $m_{\Delta F}$ is the mean logarithmic mass of $\Delta F$.

In order to reveal whether proton exerts a predominant role in the knee region of the all-particle energy spectrum, hypothesis testing is conducted \cite{methods}.
Furthermore, in order to reveal whether the CR knee is totally attributed to proton, another hypothesis testing is conducted \cite{methods}. 
If the spectral indices fitted by LHAASO are used \cite{Zhang_2024}, the significance of rejecting the presumed null hypothesis that proton makes no contribution to the knee formation is 36.8$\sigma$, while the significance of rejecting the presumed null hypothesis that the CR knee is completely ascribed to proton is 41.0$\sigma$.
In this case, the proton proportion $q_p$ that contributes to $\Delta F$ in the knee region is found to be $52.7\% \pm 1.2\% \leq q_p \leq 83.7\% \pm 0.4\%$.
Since the newly fitted results from 1 to 10~PeV yield smaller values of reduced chi-squared compared to those from LHAASO \cite{methods}, using the newly fitted results would produce more appropriate output in conducting the hypothesis test.
If the spectral indices fitted from 1 to 10~PeV are used \cite{methods}, the significance of rejecting the presumed null hypothesis that proton fails to contribute to the knee formation is 12.0$\sigma$, while the significance of rejecting the presumed null hypothesis that the knee is fully attributable to proton alone is 6.3$\sigma$. 
Under this circumstance, the proton proportion is found to be $72.6\% \pm 4.5\% \leq q_p \leq 90.6\% \pm 1.5\%$.

For one thing, the proton spectrum really bends and participates in the steepening of the all-particle energy spectrum in the knee region.
Simultaneously, proton provides the dominant contribution to $\Delta F$ around the knee energy, indicating that proton largely accounts for the formation of the CR knee.
For another, proton itself cannot completely explain the knee-like structure, suggesting that helium and even heavier nuclei may likewise contribute to $\Delta F$ in the knee region.
These findings also substantiate that proton can be accelerated to the energy exceeding the knee position, which directly supports the presence of Galactic PeV proton accelerators.

\subsection*{Rigidity-dependent knee}
\begin{figure}
           \centering
           \includegraphics[width=0.6\textwidth]{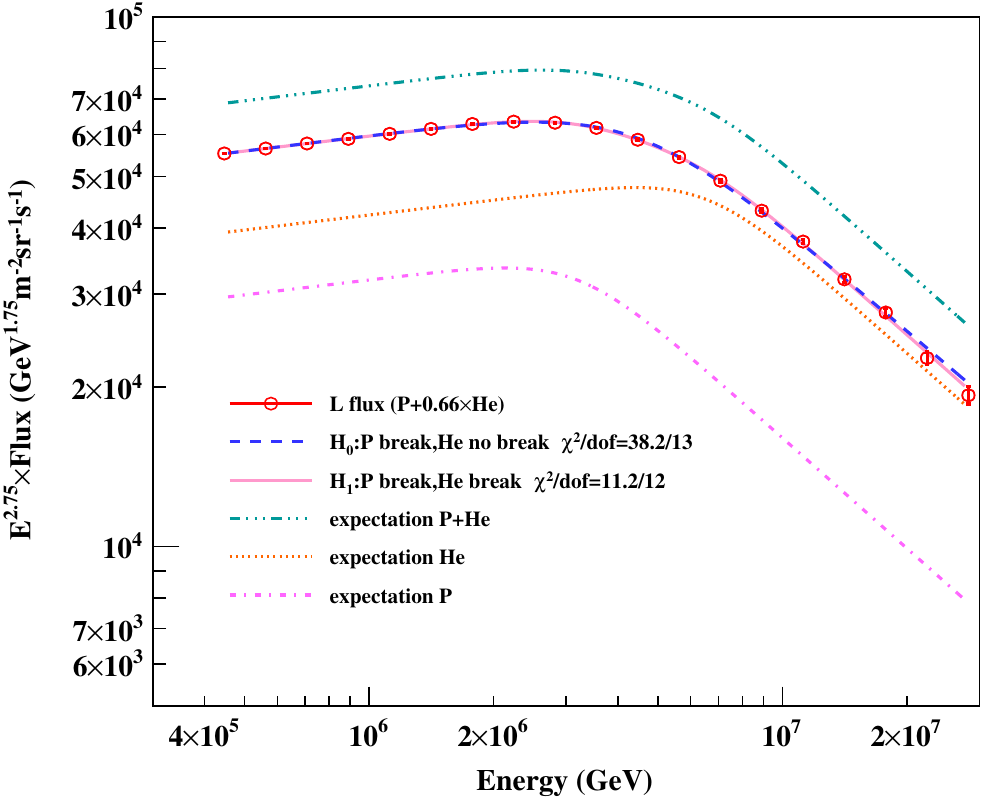}
           \caption{\textbf{The energy spectrum of light elements obtained through the cocktail strategy from 0.4 to 30~PeV.}
                         The red hollow circles represent the energy spectrum flux of light elements $L$.
                         The blue dashed line shows the fitted curve under the null hypothesis $H_0$ that only proton bends in the knee region, which yields $\chi^2 / \rm dof = 38.2/13$.
                         The pink solid line shows the fitted curve under the alternative hypothesis $H_1$ that both proton and helium bend in the knee region,  which yields $\chi^2 / \rm dof = 11.2/12$.
                         The green dash-dot-dot-dot line shows the derived energy spectrum of proton plus helium.
                         The orange dotted line shows the derived energy spectrum of helium.
                         The magenta dash-dot line shows the derived energy spectrum of proton.}
           \label{fig:L}
\end{figure}

To investigate whether the knee is  $Z$-dependent or $A$-dependent, the innovative cocktail strategy combining $F$ and $M$ is put forward. 
On the basis of $F$ and $M$, the energy spectrum flux of light elements $L$ can be created as $F - M / \ln 56$ (see Fig.~\ref{fig:L}).
Undoubtedly, the all-particle energy spectrum, the total logarithmic mass energy spectrum and the energy spectrum of light elements all act as the weighted sum of flux contributed by individual components, with the weights of different elements listed in Table~\ref{tab:weight}.
Subtracting $M / \ln 56$ from $F$ largely suppresses the contributions from heavy nuclei, and even makes iron fail to serve as a contributory factor in the energy spectrum of light elements.
There is only a minimal amount of pollution from heavy elements involved in the energy spectrum of light elements (see fig.~\ref{fig:sup_pollution}).
In that regard, $L$ almost equals to the proton flux plus 0.66 of the helium flux.
By and large, the energy spectrum of light elements is capable of characterizing the spectral properties of proton and helium.
Hence, the cocktail strategy offers a golden opportunity to construct the energy spectrum of light elements and test different cases of the CR knee.

The hypothesis test is conducted here via $L$.
In this case, the null hypothesis $H_0$ is that only one break is demonstrated in the energy spectrum of light elements, meaning that only proton bends in the knee region and the helium spectrum has no obvious structure before 30~PeV. 
While the alternative hypothesis $H_1$ is that the energy spectrum of light elements shows two breaks in the knee region, indicating that both proton and helium take part in the formation of the knee-like structure.
According to the hypothesis test \cite{methods}, the test statistic $\Delta \chi^2$ yields a p-value of $2.018 \times 10^{-7}$ (5.2$\sigma$), which signifies that both proton and helium bend in the knee region.
To reveal whether the knee is $Z$-dependent or $A$-dependent, the quantity $n$ is utilized as the cutoff energy ratio of helium to proton.
In the $Z$-dependent case, $n$ is supposed to be approximately 2; while in the $A$-dependent case, $n$ should be around 4.
Our results find that the fitted $n = 2.06 \pm 0.09$ rules out the $A$-dependent knee with 22.9$\sigma$ and is in complete agreement with the $Z$-dependent knee.
To be more specific, the cutoff energy of proton is $E_p = 3.2 \pm 0.2$~PeV and that of helium is $E_{He} = 6.6 \pm 0.5$~PeV in the knee region. 
The energy spectra of proton and helium can also be easily obtained from the fitted parameters under $H_1$, which are depicted in Fig.~\ref{fig:L}.
According to the obtained results of both proton and helium, the spectral index before the cutoff energy is $\gamma = -2.6559 \pm 0.0009$, the change in the spectral index before and beyond the cutoff position is $\Delta \gamma = -0.787 \pm 0.028$ and the sharpness parameter is $s = 5.1 \pm 0.5$ \cite{methods}.

\subsection*{Discovery of an ankle-like structure}
An ankle-like feature in the total logarithmic mass energy spectrum is clearly visible around 10~PeV.
In order to enhance the reliability of this discovery, statistical hypothesis testing is performed.
Here, the null hypothesis $H_0$ is that the total logarithmic mass energy spectrum displays a descending trend all the way from 1 to 30~PeV, and it can be described by a log-parabola function \cite{methods}.
The alternative hypothesis $H_1$ is that the total logarithmic mass energy spectrum displays an upward bend around 10~PeV, and it can be describe by a piecewise function made up of a log-parabolic function and a power-law function \cite{methods}.
The comparison between $H_0$ and $H_1$ yields a p-value of $1.031 \times 10^{-147}$ (25.9$\sigma$), thus presenting the existence of an ankle-like structure at $9.7 \pm 0.2$~PeV in the total logarithmic mass energy spectrum.
According to the fitted results under $H_1$, the spectral index before the upward bend is $-2.950 \pm 0.005$, while the spectral index after the upward bend is $-2.65 \pm 0.01$, indicating a significant hardening of the total logarithmic mass energy spectrum.

Meanwhile, the ankle-like structure in the total logarithmic mass energy spectrum is supposed to correspond to a similar structure in the all-particle energy spectrum at the same position.
In order to probe whether the all-particle energy spectrum bends around 10~PeV, another hypothesis testing is performed.
In this case, the null hypothesis $H_0$ is that the all-particle energy spectrum has no obvious structural change around 10~PeV, and it can be described by a smoothly broken power-law function.
The alternative hypothesis $H_1$ is that the structural change is noticeable at the break energy of the total logarithmic mass energy spectrum, and it can be described by a piecewise function made up of a smoothly broken power-law function and a power-law function.
The test statistic $\Delta \chi^2$ yields a p-value of $3.577 \times 10^{-9}$ (5.9$\sigma$), which indicates that the all-particle energy spectrum actually kinks around 10~PeV.

\begin{figure}
           \centering
           \includegraphics[width=0.6\textwidth]{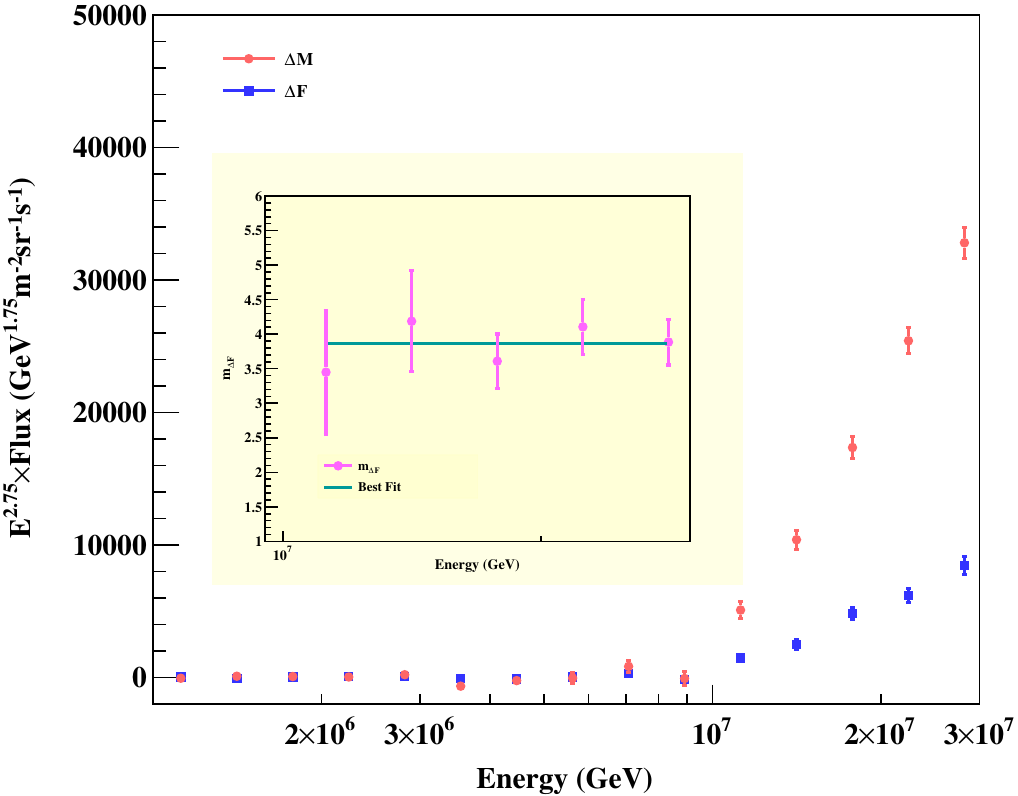}
           \caption{\textbf{An ankle-like structure stemming from the excess of iron in the total logarithmic mass energy spectrum.}
                         The surplus of the all-particle energy spectrum flux and the total logarithmic mass energy spectrum flux is shown.
                         The pink circles and the blue squares represent $\Delta M$ and $\delta F$ respectively.
                         The subfigure with a yellow background illustrates the mean logarithmic mass of the CR flux variation at each data point after 10~PeV, with the best-fit value of $m_{\Delta_F}$ ($3.9 \pm 0.2$) demonstrated by the green line.}
           \label{fig:ankle}
\end{figure}

To unravel the mystery of the ankle-like structure discovered in the total logarithmic mass energy spectrum, the mean logarithmic mass of the CR flux variation $m_{\Delta F}$ that contributes to the surplus of the all-particle energy spectrum flux $\Delta F$ beyond the knee is paid attention to.
The comparisons between measurements and extrapolations in both the all-particle energy spectrum and the total logarithmic mass energy spectrum after 10~PeV are demonstrated in fig.~\ref{fig:sup_ankleext}.
The differences, $\Delta F$ and $\Delta M$, between measurements and extrapolations can be easily calculated.
Subsequently, dividing $\Delta M$ by $\Delta F$ naturally gives $m_{\Delta F}$.
The mean logarithmic mass of $\Delta F$ at each data point after 10~PeV is listed in table~\ref{tab:sup_iron}.
If $m_{\Delta F}$ is assumed to remain unchanged after 10~PeV, as shown in Fig.~\ref{fig:ankle}, the best-fit value of $m_{\Delta F}$ is $3.9 \pm 0.2$, with the reduced chi-square statistic $\chi^2 / \rm dof = 1.2/4$.
Since the bare fact is that the logarithmic mass of aluminum (MgAlSi) is smaller than 3.332 and that of iron is approximately 4.025, the excess of the iron flux is considered as fully responsible for the ankle-like structure around 10~PeV, which implies that iron dominates the all-particle energy spectrum beyond the knee with a high concentration in CRs.

\subsection*{Summary and conclusions}
The intricate nature of the CR knee has been a subject of intense debate after its discovery.
In this paper, we propose the brand-new concept of the total logarithmic mass energy spectrum to elucidate the intrinsic characteristics of the knee.
The total logarithmic mass energy spectrum is a weighted spectrum where each component is weighted by its logarithmic mass, with heavier nuclei assigned greater weights. 
Compared to the all-particle energy spectrum, it is more sensitive to the spectral structures of heavier nuclei.
Through combining the total logarithmic mass energy spectrum and the all-particle energy spectrum, we find that proton constitutes the dominant contribution to the formation of the CR knee. 
Furthermore, an ankle-like structure is discovered at $9.7 \pm 0.2$~PeV, featuring a pronounced upward bend, which emanates from the excess of the iron spectrum.

Mixing the all-particle energy spectrum flux and the total logarithmic mass energy spectrum flux magically, we obtained the energy spectrum of light elements weighted by $1 - \ln A / \ln 56$. 
Through the cocktail strategy, the case of a mass-dependent knee is ruled out with a significance of 22.9$\sigma$, suggesting that the cutoff energies of different species exhibit rigidity-dependence.
Our findings reveal that the knee position in the proton spectrum is at $3.2 \pm 0.2$~PeV, and that in the helium spectrum occurs at $6.6 \pm 0.5$~PeV.
The unveiling of the rigidity-dependent knee for the first time marks a pivotal moment in resolving the long-standing puzzle on the origins of CRs.

\newpage

\begin{table}
          \centering
          \scriptsize     
          \caption{\textbf{Weights of different elements in the three weighted energy spectra.}
                        For the all-particle energy spectrum, the weight factors of all the elements are equally 1. 
                        For the total logarithmic mass energy spectrum, each component is weighted by its logarithmic mass $\ln A$. 
                        For the energy spectrum of light elements, each component is weighted by $1 - \ln A / \ln 56$.}         
          \label{tab:weight} 
          \begin{tabular}{ccccccc} 
          \\
          \hline
           Energy spectrum    &   Weight factor        &   Proton   &   Helium   &   Nitrogen  &   Aluminum  &   Iron       \\
          \hline
	  $F$                          &   $1$                       &   $1$        &   $1$        &   $1$         &   $1$            &   $1$        \\
	  $M$                         &   $\ln A$                  &   $0$        &   $1.39$   &   $2.64$    &   $3.30$       &   $4.03$   \\
	  $L$                          &   $1 - \ln A/\ln 56$   &   $1$        &   $0.66$   &   $0.34$    &   $0.18$       &   $0$        \\
	 \hline
         \end{tabular}
\end{table}

\clearpage

\bibliography{knee.bib}
\bibliographystyle{sciencemag}

\section*{Acknowledgments}
\paragraph*{Funding:}
H.~H., H.~Z., Q.~C., L.~M., and C.~F. were supported by the National Natural Science Foundation of China under grants 12275280, 12405131, 12175121 and 12261160362.

\paragraph*{Author contributions:}
H.~H. initiated the study, proposed the concept of total logarithmic mass energy spectrum and led the first discoveries.
H.~Z. helped construct the initial sample, performed data analysis and contributed to interpretation.
Q.~C. (supervised by H.~H.) consolidated the results and authored most of the text.
L.~M. coordinated the entire data analysis, participated in the interpretation of results and contributed to manuscript structure.
C.~F. contributed to the overall interpretation of the results and various aspects of the analysis.

\paragraph*{Competing interests:}
There are no competing interests to declare.

\paragraph*{Data and materials availability:}
The LHAASO measurement results were obtained from the National HEP Data Center at https://www.nhepsdc.cn/resource/astro/lhaaso/paper-PRL2024-LL18217.

\subsection*{Supplementary materials}
Materials and Methods\\
Figs. S1 to S5\\
Tables S1 to S6\\
References \textit{(24-\arabic{enumiv})}

\newpage

\renewcommand{\thefigure}{S\arabic{figure}}
\renewcommand{\thetable}{S\arabic{table}}
\renewcommand{\theequation}{S\arabic{equation}}
\renewcommand{\thepage}{S\arabic{page}}
\setcounter{figure}{0}
\setcounter{table}{0}
\setcounter{equation}{0}
\setcounter{page}{1}

\begin{center}
\section*{Supplementary Materials for\\ \scititle}
               Huihai~He,
               Hengying~Zhang,
               Qinyi~Cheng,
               Lingling~Ma$^{\ast}$,
               Cunfeng~Feng\\ 
               \small$^\ast$Corresponding author. Email: llma@ihep.ac.cn
\end{center}

\subsubsection*{This PDF file includes:}
                          Materials and Methods\\
                          Figures S1 to S5\\
                          Tables S1 to S6\\

\newpage

\subsection*{Materials and Methods}

\subsubsection*{Spectrum fitting from 1 to 10~PeV}
In light of considerably large reduced chi-squared presented in the fit of energy spectrum published by LHAASO \cite{Zhang_2024}, the 10 data points from 1 to 10~PeV are selected to fit the all-particle energy spectrum and the mean logarithmic mass via a smoothly broken power-law function, which can be expressed as
\begin{equation}
           J(E) = J_0 \left(\frac{E}{1~\rm PeV}\right)^{\gamma} [1+\left(\frac{E}{E_b}\right)^s]^{\Delta \gamma /s},
          \label{eq:sup_SBPL}
\end{equation}
where $E_b$ corresponds to the transition position, $\gamma$ is the spectral index before the transition, $\Delta \gamma$ is the spectral change before and beyond the transition, $s$ is the sharpness parameter of the transition. 
Fitted parameters of the all-particle energy spectrum and the mean logarithmic mass in the knee region are listed in table~\ref{tab:sup_FF} and table~\ref{tab:sup_FA} respectively, with fitted curves demonstrated in Fig.~\ref{fig:sup_newfit}.

According to the newly fitted results from 1 to 10~PeV, for the all-particle energy spectrum, the knee position is at $4.40 \pm 0.32$~PeV, which is located after the published knee position $3.67 \pm 0.05$~PeV from LHAASO \cite{Zhang_2024}.
For the mean logarithmic mass, the transition position is at $4.21 \pm 0.25$~PeV, which is close to the updated knee position in the all-particle energy spectrum.
Compared to the fitted results reported by LHAASO \cite{Zhang_2024}, the newly fitted results of both the all-particle energy spectrum and the mean logarithmic mass yield surprisingly small values of reduced chi-squared.

\subsubsection*{Hypothesis testing for proton knee}
\begin{figure}
          \centering
          \includegraphics[width=0.6\textwidth]{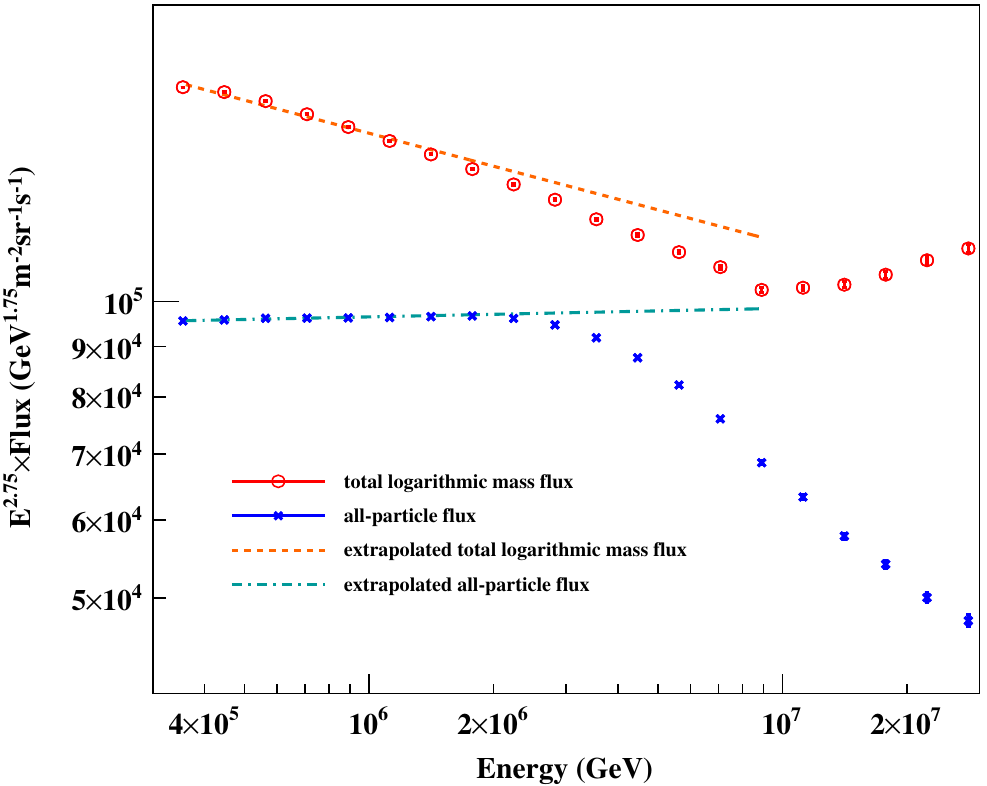} 
          \caption{\textbf{Comparisons between measurements and extrapolations in the knee region.}
                        The red hollow circles represent the total logarithmic mass energy spectrum flux $M$, while the blue crosses represent the all-particle energy spectrum flux $F$. 
                        The orange dashed line and the green dash-dot line show the extrapolated $M'$ and the extrapolated $F'$ respectively.
                        The spectral indices before the knee from \cite{Zhang_2024} are used to obtain the extrapolated values of flux.}
          \label{fig:sup_prlext}
\end{figure}

\begin{figure}
           \centering
           \includegraphics[width=0.6\textwidth]{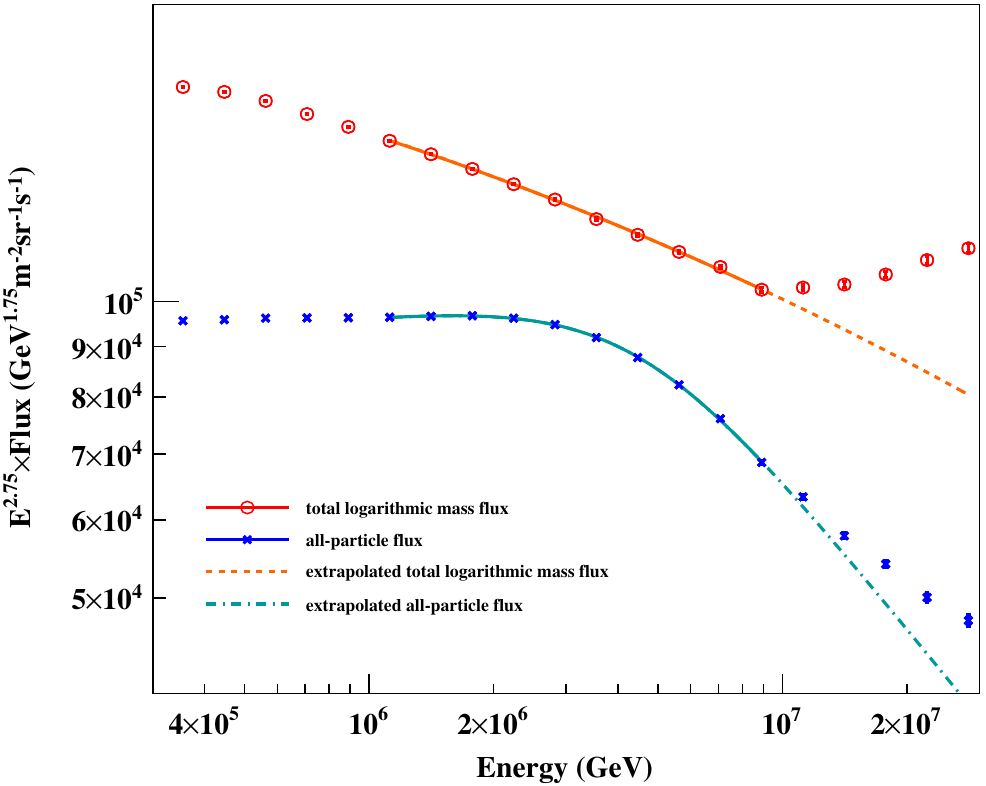}
           \caption{\textbf{Comparisons between measurements and extrapolations after 10~PeV.}
                         Same as Fig.~\ref{fig:sup_prlext}.}
           \label{fig:sup_ankleext}
\end{figure}

On the assumption that the falling trends of the all-particle energy spectrum, the mean logarithmic mass of CRs and the total logarithmic mass energy spectrum remain unchanged in the knee region, all of them can be described by a single power-law function.
In this case, the extrapolated flux $F'$ of the all-particle energy spectrum can be expressed as
\begin{equation}
           F' = F_0 \left(\frac{E}{1~\rm PeV}\right)^{\gamma_F},
          \label{eq:sup_extF}
\end{equation}
where $F_0$ is the all-particle energy spectrum flux at 1~PeV and $\gamma_F$ is the spectral index before the knee.
According to the reported measurement results from LHAASO, the spectral index of the all-particle energy spectrum is $-2.7413 \pm 0.0004$ before the knee \cite{Zhang_2024}.
The comparison between the measured flux $F$ and the extrapolated flux $F'$ of the all-particle energy spectrum in the knee region is shown in Fig.~\ref{fig:sup_prlext}.
The difference $\Delta F$ between measurements and extrapolations represents the loss of the all-particle energy spectrum flux caused by the rapid drop of CR spectra for individual species around the knee.
Similarly, the extrapolated mean logarithmic mass $m'$ can be expressed as
\begin{equation}
           m' = m_0 \left(\frac{E}{1~\rm PeV}\right)^{\gamma_m},
          \label{eq:sup_extm}
\end{equation}
where $m_0$ is the mean logarithmic mass of CRs at 1~PeV, and $\gamma_m$ is the spectral index before the transition energy.
The spectral index of the mean logarithmic mass published by LHAASO is $-0.1200 \pm 0.0003$ before the transition of CR components from light elements on average to heavy elements on average \cite{Zhang_2024}.
Based on $F'$ and $m'$, the extrapolated flux $M'$ of the total logarithmic mass energy spectrum can be expressed as
\begin{equation}
           M' = F' \times m' = F_0 \times m_0 \left(\frac{E}{1~\rm PeV}\right)^{\gamma_F + \gamma_m}.
          \label{eq:sup_extM}
\end{equation}
As depicted in Fig.~\ref{fig:sup_prlext}, the difference $\Delta M$ between measurements and extrapolations represents the loss of the total logarithmic mass energy spectrum flux caused by the CR nuclei except for proton.

Combining the calculated $\Delta F$ and $\Delta M$, the mean logarithmic mass of the CR flux variation $m_{\Delta F}$ can be derived using the following formula:
\begin{equation}
           m_{\Delta F} = \frac{\Delta M}{\Delta F}.
          \label{eq:sup_losslnA}
\end{equation}
When the mass composition of CRs is taken into consideration, $m_{\Delta F}$ can also be expressed as 
\begin{equation}
           m_{\Delta F} = q_p \times m_p + (1-q_p) \times m_{\geq He}= (1-q_p) \times m_{\geq He}, 
          \label{eq:sup_lossm}
\end{equation}
where $q_p$ is the proton proportion that contributes to $\Delta F$ in the knee region, $m_p$ (=0) is the logarithmic mass of proton, $m_{\geq He}$ ($\geq \ln 4$) is the mean logarithmic mass of helium and heavier nuclei that contribute to $\Delta F$ in the knee region and $m_{\Delta F}$ is hereafter referred to as $X$.
So the expected flux $M''$ of the total logarithmic mass energy spectrum can be expressed as
\begin{equation}
           M'' = M' - \Delta F \times X.
          \label{eq:sup_expM}
\end{equation}

In order to investigate whether proton plays a dominant role in the formation of the CR knee, hypothesis testing \cite{Wilks_1938} is implemented. 
Under this circumstance, the null hypothesis $H_0$ is that proton shows no visible sign of cutoff in the energy spectrum,
thus making no contribution to the spectral break in the knee region with $q_p = 0$; while the alternative hypothesis $H_1$ is that proton bends and contributes to $\Delta F$ in the knee region with $q_p > 0$.
To avoid the energy-dependence of the proton proportion in the following statistical tests, the cumulative sum of all the calculated quantities is taken advantage of here.
The calculated cumulative sum of lost flux in the all-particle energy spectrum is $\Delta F=\sum_i \delta F_i$ with its corresponding error $\sigma_{\Delta F}$, while that of flux in the total logarithmic mass energy spectrum is $M=\sum_i M_i$ with its corresponding error $\sigma_{M}$.
In extrapolation, the cumulative sum of the total logarithmic mass energy spectrum flux is $M'=\sum_i M'_i$ with its corresponding error $\sigma_{M'}$.
The expected cumulative sum of the total logarithmic mass energy spectrum flux is $M'' = M' - \Delta F \times X$ with its corresponding error $\sigma_{M''
}=\sqrt{\sigma_{M'}^2 + (\sigma_{\Delta F} \times X)^2}$.
Since the measured flux follows Gaussian distribution, the likelihood function can be expressed as
\begin{equation}
           \mathcal L = \frac{1}{\sqrt{2\pi} \sigma} e^{-\frac{(M - M'')^2}{2 \sigma^2}},
          \label{eq:sup_likeli}
\end{equation}
where the total error is $\sigma = \sqrt{\sigma_{M}^2 + \sigma_{M''}^2}$.
Accordingly, the log-likelihood function is 
\begin{equation}
           \ln \mathcal L = -\frac{(M - M'')^2}{2\sigma^2} - \ln \sigma.
          \label{eq:sup_lnlikeli}
\end{equation}
In this case, the test statistic can be denoted as 
\begin{equation}
           TS = 2(\ln{\mathcal L(X_1)}-\ln{\mathcal L(X_0)}),
          \label{eq:sup_TS}
\end{equation}
where $X_0$ is the best-fit parameter under $H_0$ which maximizes $\mathcal L$ when $q_p = 0$ and $X_1$ is the best-fit parameter under $H_1$.
The significance here is just $\sqrt{TS}$.

It is important to note that only the 4 measured points after the knee energy with $\delta F_i / \sigma_{\delta F_i} > 5$ are utilized in selection to prevent interference from fluctuations of the all-particle energy spectrum flux.
If the spectral indices published by LHAASO are used, the best-fit $X_1$ is $0.656 \pm 0.017$ with $X_0 = \ln 4$.
The significance of rejecting $H_0$ when $H_1$ is true is 36.8$\sigma$, indicating that proton really bends and contributes to $\Delta F$ in the knee region.
For one thing, $m_{\geq He} \geq \ln 4$ yields $q_p \geq 52.7\% \pm 1.2\%$, which gives the lower limit of the proton proportion.
For another, iron has the largest logarithmic mass of $\ln 56$, so $m_{\geq He}$ must be equal to or smaller than $\ln 56$, thus yeilding $q_p \leq 83.7\% \pm 0.4\%$.

Then, in order to investigate whether the CR knee is totally attributed to proton, another hypothesis testing is implemented here. 
Under this circumstance, the null hypothesis $H_0$ is that the CR knee is completely ascribed to proton with the proton proportion $q_p = 1$, while the alternative hypothesis $H_1$ is that proton cannot suffice to explain the knee entirely with the proton proportion $q_p < 1$.
In this case, the best-fit $X_1$ is still $0.656 \pm 0.017$ with $X_0 = 0$, and the corresponding significance of rejecting $H_0$ when $H_1$ is true is 41.0$\sigma$.
Consequently, the knee formation cannot be fully attributable to proton alone.

Since the newly fitted results from 1 to 10~PeV yield smaller values of reduced chi-squared in contrast to those from LHAASO \cite{Zhang_2024}, using the newly fitted results can produce more appropriate output in performing the hypothesis test.
If the spectral indices of the newly fitted results are used, the best-fit $X_1$ is $0.380 \pm 0.062$.
In this case, the significance of rejecting the presumed null hypothesis that proton makes no contribution to the knee formation is 12.0$\sigma$, while the significance of rejecting the presumed null hypothesis that the knee is totally attributed to proton is 6.3$\sigma$.
Simultaneously, $\ln 4 \leq m_{\geq He} \leq \ln 56$ yields $72.6\% \pm 4.5\% \leq q_p \leq 90.6\% \pm 1.5\%$.

\subsubsection*{Hypothesis testing for rigidity-dependent knee}
In order to investigate whether the knee is  $Z$-dependent or $A$-dependent, hypothesis testing is implemented here via $L$.
In this case, the null hypothesis $H_0$ is that only one break is demonstrated in the energy spectrum of light elements, meaning that only proton bends in the knee region and the helium spectrum represented by the single power-law distribution has no obvious structure before 30~PeV. 
$H_0$ can be denoted as
\begin{equation}
           L(E) = L_0 \left(\frac{E}{1~\rm PeV}\right)^{\gamma} \{f_p[1+\left(\frac{E}{E_p}\right)^s]^{\Delta \gamma /s} + (1-\frac{\ln 4}{\ln 56})(1-f_p)\},
          \label{eq:sup_Lone}
\end{equation}
where $f_p$ is the proton ratio in the energy spectrum of light elements and $E_p$ is the cutoff energy of proton.
While the alternative hypothesis $H_1$ is that the energy spectrum of light elements shows two breaks in the knee region, indicating that both proton and helium take part in the formation of the knee-like structure.
So $H_1$ can be denoted as
\begin{equation}
           L(E) = L_0 \left(\frac{E}{1~\rm PeV}\right)^{\gamma} \{f_p[1+\left(\frac{E}{E_p}\right)^s]^{\Delta \gamma /s} + (1-\frac{\ln 4}{\ln 56})(1-f_p)[1+\left(\frac{E}{n E_p}\right)^s]^{\Delta \gamma /s}\},
          \label{eq:sup_Ltwo}
\end{equation}
where $n$ is the quantity to reveal whether the knee is $Z$-dependent or $A$-dependent.
In the $Z$-dependent case, $n$ is supposed to be approximately 2; while in the $A$-dependent case, $n$ should be around 4.

Owing to the large deviation of the first point observed in the total logarithmic mass energy spectrum, it is not used in the following analysis.
The 19 data points from 0.4 to 30~PeV are utilized to fit the energy spectrum of light elements.
The fitted parameters of the energy spectrum of light elements from 0.4 to 30~PeV are listed in Table~\ref{tab:sup_L}.
The significance of rejecting the presumed null hypothesis that only one break is demonstrated in the energy spectrum of light elements is 5.2$\sigma$, suggesting the sequential cutoff of proton and helium.
Subsequently, the energy spectra of proton and helium can also be easily obtained from the energy spectrum of light elements.
The proton spectrum can be expressed as 
\begin{equation}
           F_p(E) = L_0 \left(\frac{E}{1~\rm PeV}\right)^{\gamma} f_p[1+\left(\frac{E}{E_p}\right)^s]^{\Delta \gamma /s}.
          \label{eq:sup_Fp}
\end{equation}
The helium spectrum can be expressed as 
\begin{equation}
           F_{He}(E) = L_0 \left(\frac{E}{1~\rm PeV}\right)^{\gamma} (1-f_p)[1+\left(\frac{E}{n E_p}\right)^s]^{\Delta \gamma /s}.
          \label{eq:sup_FHe}
\end{equation}

\subsubsection*{Hypothesis testing for the ankle-like structure}
\begin{figure}
           \centering
           \includegraphics[width=0.6\textwidth]{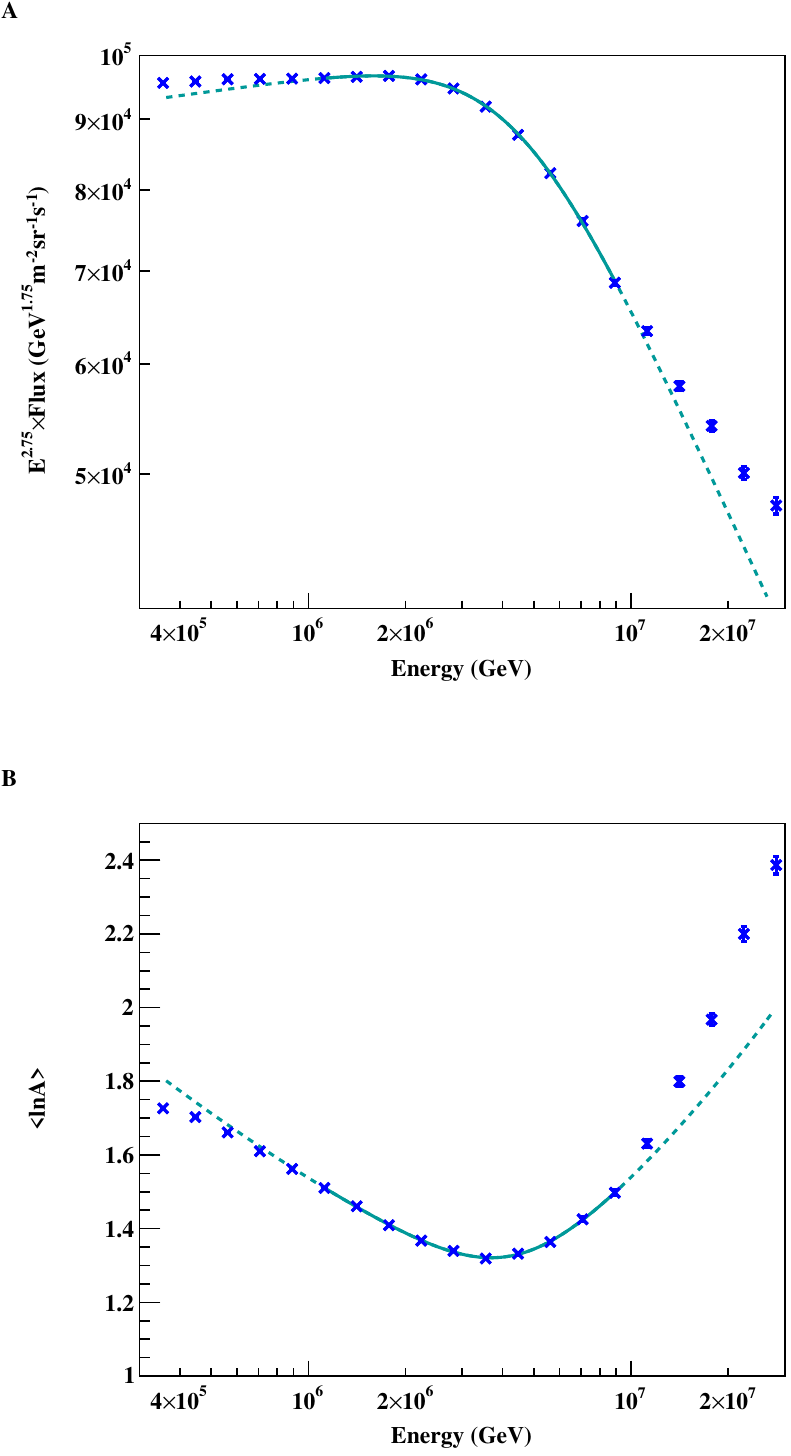}
           \caption{\textbf{Fitted results of energy spectrum in the knee region when the 10 data points from 1 to 10~PeV are taken into consideration.}
                        (\textbf{A}) The all-particle energy spectrum flux multiplied by E$^{2.75}$ as a function of energy is fitted using a smoothly broken power-law function.
                        (\textbf{B}) The mean logarithmic mass of CRs as a function of energy is also fitted using a smoothly broken power-law function.}
           \label{fig:sup_newfit}
\end{figure}

\begin{figure}
           \centering
           \includegraphics[width=0.6\textwidth]{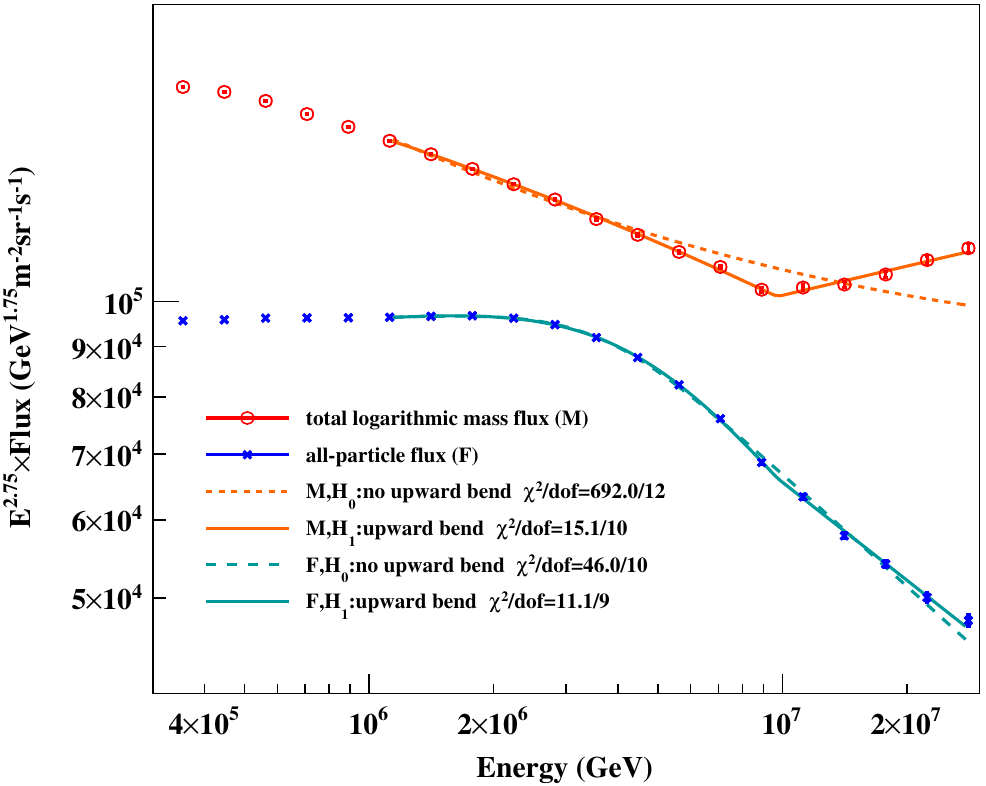}
           \caption{\textbf{Fitted results of the energy spectra from 1 to 30~PeV are illustrated to present enough evidence of the ankle-like structure.}
                         The red hollow circles represent the total logarithmic mass energy spectrum flux $M$, while the blue crosses represent the all-particle energy spectrum flux $F$. 
                         The orange dashed line shows the fitted curve of the total logarithmic mass energy spectrum under the null hypothesis $H_0$ that no ankle-like structure is observed, which yields $\chi^2 / \rm dof = 692.0/12$.
                         The orange solid line shows the fitted curve of the total logarithmic mass energy spectrum under the alternative hypothesis $H_1$ that an ankle-like structure is observed, which yields $\chi^2 / \rm dof = 15.1/10$.
                         The green dashed line shows the fitted curve of the all-particle energy spectrum under the null hypothesis $H_0$ that no structural change is observed, which yields $\chi^2 / \rm dof = 46.0/10$.
                         The green solid line shows the fitted curve of the all-particle energy spectrum under the alternative hypothesis $H_1$ that the structural change is noticeable, which yields $\chi^2 / \rm dof = 11.1/9$.}
           \label{fig:sup_ankle}
\end{figure}

To enhance the reliability of the ankle-like structure in the total logarithmic mass energy spectrum, hypothesis testing is implemented.
Given relatively large deviations of the 5 data points below 1~PeV observed in both the all-particle energy spectrum and the mean logarithmic mass (see Fig.~\ref{fig:sup_newfit}), they are excluded from the following analysis.
Here, the null hypothesis $H_0$ is that the total logarithmic mass energy spectrum from 1 to 30~PeV can be well fitted by a log-parabola function, which is defined by the following equation:
\begin{equation}
           M(E) = M_0 \left( \frac{E}{1~\rm PeV} \right)^{-\alpha -\beta \log \left( \frac{E}{1~\rm PeV} \right)},
          \label{eq:sup_Mlog}
\end{equation}
where $\log$ refers to the natural logarithm, $M_0$ is the total logarithmic mass energy spectrum flux at 1~PeV, $\alpha$ is the spectral index and $\beta$ is the curvature parameter. 
The alternative hypothesis $H_1$ is that the total logarithmic mass energy spectrum from 1 to 30~PeV can be well fitted by a piecewise function made up of a log-parabolic function and a power-law function, which is defined by the following equation:
\begin{equation}
           M(E) =
           \begin{cases}
           M_0 \left( \frac{E}{E_b} \right)^{-\alpha -\beta \log \left( \frac{E}{E_b} \right)}, & E < E_b \\
           M_0 \left( \frac{E}{E_b} \right)^{\gamma}, & E \geq E_b \\
           \end{cases}
          \label{eq:sup_Mlogpl}
\end{equation}
where $E_b$ is the break energy, $M_0$ is the total logarithmic mass energy spectrum flux at the break energy and $\gamma$ is the index of the power-law function.
The fitted parameters of the total logarithmic mass energy spectrum from 1 to 30~PeV are listed in Table~\ref{tab:sup_tlmankle}, with fitted curves depicted in Fig.~\ref{fig:sup_ankle}.

To investigate whether the all-particle energy spectrum bends around 10~PeV, another hypothesis testing is conducted.
In this case, the null hypothesis $H_0$ is that the all-particle energy spectrum from 1 to 30~PeV can be well fitted by a smoothly broken power-law function.
The alternative hypothesis $H_1$ is that the structural change is noticeable at the break energy of the total logarithmic mass energy spectrum.
In other words, the all-particle energy spectrum from 1 to 30~PeV can be well fitted by a piecewise function made up of a smoothly broken power-law function and a power-law function, which is defined by the following equation:
\begin{equation}
           F(E) =
           \begin{cases}
           F_0 \left(\frac{E}{1~\rm PeV}\right)^{\gamma} [1+\left(\frac{E}{E_{\rm cut}}\right)^s]^{\Delta \gamma / s}, & E < E_b \\
           F_0 \left(\frac{E_b}{1~\rm PeV}\right)^{\gamma} [1+\left(\frac{E_b}{E_{\rm cut}}\right)^s]^{\Delta \gamma / s} \times \left( \frac{E}{E_b} \right)^{\gamma'}, & E \geq E_b \\
           \end{cases}
          \label{eq:sup_Fsbplpl}
\end{equation}
where $\gamma$ is the spectral index before the knee, $E_{\rm cut}$ is the knee energy observed in the all-particle energy spectrum, $\Delta \gamma$ is the spectral change, $s$ is the sharpness parameter of the knee, $E_b$ is the fitted break energy in the total logarithmic mass spectrum and $\gamma'$ is the index of the power-law function.
The fitted parameters of the all-particle energy spectrum from 1 to 30~PeV are listed in Table~\ref{tab:sup_Fankle}, with fitted curves depicted in Fig.~\ref{fig:sup_ankle}.

\subsubsection*{Pollution of heavy elements in $L$}
\begin{figure}
           \centering
           \includegraphics[width=0.6\textwidth]{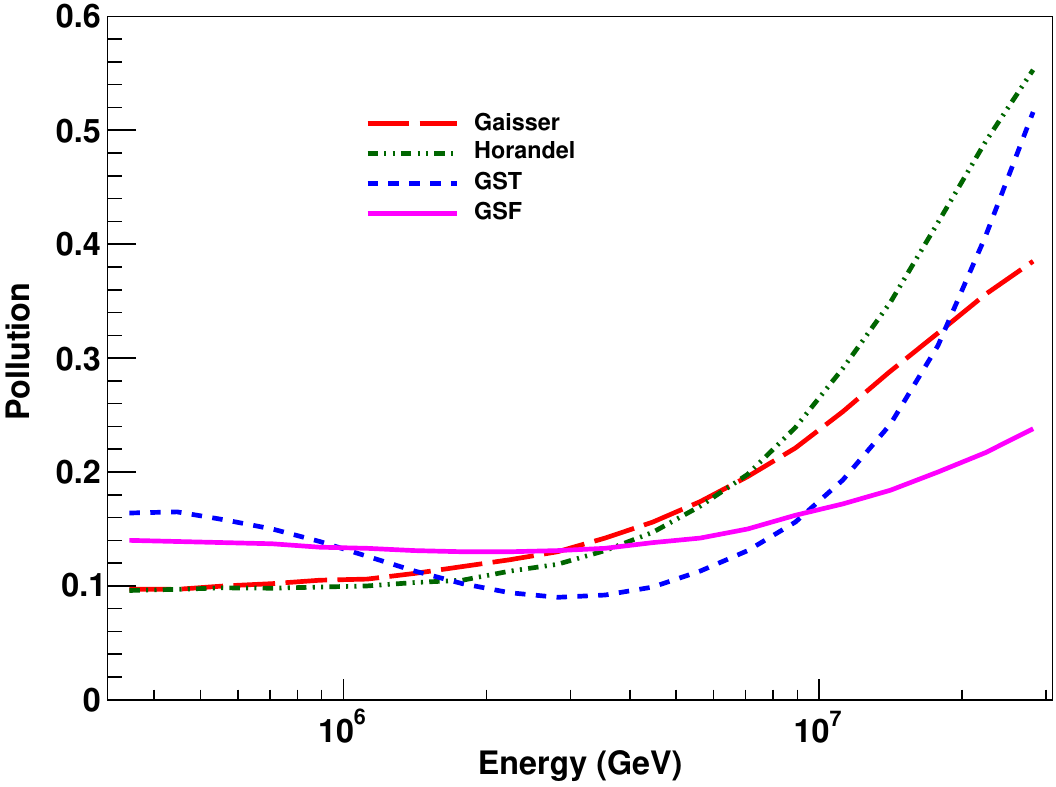}
           \caption{\textbf{The pollution of heavy elements in the energy spectrum of light elements calculated under different composition models.}
                         The red long dashed line represents the pollution calculated under the Gaisser model \cite{Gaisser_2013}.
                         The green dash-dot-dot line represents the pollution calculated under the Horandel model \cite{Horandel_2003}.
                         The blue short dashed line represents the pollution calculated under the GST model \cite{Stanev_2014}.
                         The magenta solid line represents the pollution calculated under the GSF model \cite{Dembinski_2017}.}
           \label{fig:sup_pollution}
\end{figure}

To present evidence of the plausibility of regarding the energy spectrum of light elements as the energy spectrum of proton weighted by 1 plus He weighted by 0.66, the pollution of heavy elements is discussed in detail.
The ratios of individual energy spectrum to the all-particle energy spectrum at different energies are taken into consideration.
The weight factor $1 - \ln A / \ln 56$ multiplied by the ratio of individual species makes the weight ratio of every individual energy spectrum.
Then, the pollution of heavy elements is the sum of the weight ratio of nitrogen (CNO), aluminum (MgAlSi) and iron (Fe).
The pollution of heavy elements calculated under different composition models (such as Gaisser \cite{Gaisser_2013}, Horandel \cite{Horandel_2003}, GST \cite{Stanev_2014}, and GSF \cite{Dembinski_2017}) is demonstrated in Fig~\ref{fig:sup_pollution}.
Obviously, the pollution of heavy elements is below 15\% in the knee region under the GST \cite{Stanev_2014} and GSF \cite{Dembinski_2017} models.
The pollution rapidly increases beyond the knee, even reaching up to 50\% around 30~PeV under the Gaisser \cite{Gaisser_2013} and Horandel \cite{Horandel_2003} models.
As a matter of fact, the mean logarithmic mass $\langle \ln A \rangle$ in the composition models cannot match the measured results of LHAASO except for the GSF model.
Most of these models indicate a much heavier $\langle \ln A \rangle$ than the measured results.
Specifically speaking, the measured $\langle \ln A \rangle$ is close to the logarithmic mass of helium below 10~PeV, suggesting a dominance of light elements in the CRs. 
However, these models assume that the mean logarithmic mass is close to the logarithmic mass of nitrogen at 10~PeV, which overestimates the concentration of nitrogen.
Therefore, the pollution in the energy spectrum of light elements is mostly attributed to nitrogen.
Undoubtedly, adopting these models significantly aggravates the pollution of heavy elements.

\newpage

\begin{table}
          \centering      
          \scriptsize  
          \caption{\textbf{Fitted parameters of the all-particle energy spectrum in the knee region.}
                        The all-particle energy spectrum flux as a function of energy is fitted using a smoothly broken power-law function.
                        The upper row shows the fitted results when all of the 20 data points from 0.3 to 30~PeV are taken into consideration.
                        The lower row shows the fitted results when the 10 data points from 1 to 10~PeV are taken into consideration.}
          \label{tab:sup_FF} 
          \begin{tabular}{ccccccc} 
          \\
          \hline
          Energy region   &   $F_0$                                                                                &   $\gamma$                      &   $\Delta \gamma$       &   $E_b$                   &   $s$                   &   $\chi^2 / \rm dof$   \\
                                   &   (10$^{-12}$~GeV$^{-1}$\,m$^{-2}$\,sr$^{-1}$\,s$^{-1}$)   &                                          &                                     &   (PeV)                    &                            &                                 \\
	 \hline
	  0.3-30~PeV     &   $3.0509 \pm 0.0009$                                                          &   $-2.7413 \pm 0.0004$   &   $-0.387 \pm 0.005$   &   $3.67 \pm 0.05$   &   $4.2 \pm 0.1$   &   $94.1/15$              \\
	  1-10~PeV        &   $3.0492 \pm 0.0041$                                                          &   $-2.7176 \pm 0.0075$   &   $-0.538 \pm 0.056$   &   $4.40 \pm 0.32$   &   $2.7 \pm 0.3$   &   $5.8/5$                  \\
	 \hline
         \end{tabular}
\end{table}

\begin{table}
          \centering
          \scriptsize      
          \caption{\textbf{Fitted parameters of the mean logarithmic mass in the knee region.}
                        Similar in form to Table~\ref{tab:sup_FF}.}
          \label{tab:sup_FA} 
          \begin{tabular}{ccccccc} 
          \\
          \hline
           Energy region   &   $m_0$                          &   $\gamma$                     &   $\Delta \gamma$      &   $E_b$~(PeV)       &   $s$                   &   $\chi^2 / \rm dof$   \\
          \hline
	  0.3-30~PeV       &   $1.5361 \pm 0.0004$   &   $-0.1200 \pm 0.0003$   &   $0.497 \pm 0.008$   &   $5.36 \pm 0.08$   &   $5.7 \pm 0.3$   &   $2455.6/15$          \\
	  1-10~PeV          &   $1.5371 \pm 0.0014$   &   $-0.1564 \pm 0.0053$   &   $0.413 \pm 0.040$   &   $4.21 \pm 0.25$   &   $3.7 \pm 0.5$   &   $5.0/5$                  \\
	 \hline
         \end{tabular}
\end{table}

\begin{table}
          \centering 
          \tiny      
          \caption{\textbf{Fitted parameters of the energy spectrum of light elements from 0.4 to 30~PeV.}
                        The null hypothesis $H_0$ means that only proton bends in the knee region.
                        The alternative hypothesis $H_1$ means that both proton and helium bend in the knee region.}
          \label{tab:sup_L}
          \begin{tabular}{ccccccccc} 
          \\
          \hline
           Hypothesis   &   $L_0$                                                                                 &   $\gamma$                     &   $\Delta \gamma$      &   $E_p$                   &   $f_p$ @ 1~PeV    &   $s$                   &   $n$                           &   $\chi^2 / \rm dof$   \\
                                &   (10$^{-12}$~GeV$^{-1}$\,m$^{-2}$\,sr$^{-1}$\,s$^{-1}$)   &                                         &                                     &   (PeV)                    &                                &                            &                                    &                                 \\
	 \hline
	  $H_0$           &   $1.8893 \pm 0.0012$                                                         &   $-2.6528 \pm 0.0010$   &   $-0.762 \pm 0.017$   &   $4.49 \pm 0.09$   &   $1.00 \pm 0.01$   &   $3.3 \pm 0.1$   &                                    &   $38.2/13$             \\
	  $H_1$           &   $2.3457 \pm 0.0602$                                                         &   $-2.6559 \pm 0.0009$   &   $-0.787 \pm 0.028$   &   $3.18 \pm 0.21$   &   $0.43 \pm 0.06$   &   $5.1 \pm 0.5$   &   $2.055 \pm 0.086$   &   $11.2/12$              \\
	 \hline
         \end{tabular}
\end{table}

\begin{table}
          \centering          
          \scriptsize
          \caption{\textbf{Fitted parameters of the total logarithmic mass energy spectrum from 1 to 30~PeV.}
                        The null hypothesis $H_0$ means that the total logarithmic mass energy spectrum has no ankle-like structure around 10~PeV.
                        The alternative hypothesis $H_1$ means that the total logarithmic mass energy spectrum demonstrates an ankle-like structure.}
          \label{tab:sup_tlmankle}
          \begin{tabular}{ccccccc} 
          \\
          \hline
           Hypothesis   &   $M_0$                                                                          &   $\alpha$                   &   $\beta$                      &   $E_b$                   &   $\gamma$                 &   $\chi^2 / \rm dof$   \\
                                &    (GeV$^{-1}$\,m$^{-2}$\,sr$^{-1}$\,s$^{-1}$)               &                                    &                                     &   (PeV)                    &                                     &                                  \\
	 \hline
	  $H_0$           &   $(4.725 \pm 0.004) \times 10^{-12}$ @ 1~PeV           &   $2.935 \pm 0.002$   &   $-0.018 \pm 0.001$   &                                &                                     &   $692.0/12$             \\
	  $H_1$           &   $(6.175 \pm 0.028) \times 10^{-15}$ @ $E_b$~PeV   &   $2.950 \pm 0.005$   &   $0.014 \pm 0.002$    &   $9.70 \pm 0.24$   &   $-2.652 \pm 0.011$   &   $15.1/10$               \\
	 \hline
         \end{tabular}
\end{table}

\begin{table}
          \centering          
          \scriptsize
          \caption{\textbf{Fitted parameters of the all-particle energy spectrum from 1 to 30~PeV.}
                        The null hypothesis $H_0$ means that the all-particle energy spectrum has no structure around 10~PeV.
                        The alternative hypothesis $H_1$ means that the all-particle energy spectrum bends around 10~PeV.}
          \label{tab:sup_Fankle}
          \begin{tabular}{cccccccc} 
          \\
          \hline
           Hypothesis   &   $F_0$                                                                                 &   $\gamma$                 &   $\Delta \gamma$       &   $E_{\rm cut}$       &   $s$                    &   $\gamma'$               &   $\chi^2 / \rm dof$   \\
                                &   (10$^{-12}$~GeV$^{-1}$\,m$^{-2}$\,sr$^{-1}$\,s$^{-1}$)   &                                     &                                     &   (PeV)                    &                             &                                    &                                 \\
	 \hline
	  $H_0$           &   $3.044 \pm 0.002$                                                             &   $-2.734 \pm 0.004$   &   $-0.396 \pm 0.008$   &   $3.63 \pm 0.05$   &   $3.9 \pm 0.2$   &                                     &   $46.0/10$             \\
	  $H_1$           &   $3.050 \pm 0.004$                                                             &   $-2.717 \pm 0.007$   &   $-0.552 \pm 0.051$   &   $4.47 \pm 0.29$   &   $2.7 \pm 0.3$   &   $-3.080 \pm 0.010$   &  $11.1/9$                \\
	 \hline
         \end{tabular}
\end{table}

\begin{table}
          \centering          
          \caption{\textbf{The excess of the mean logarithmic mass at each data point after 10~PeV.}
                       $\Delta M / \Delta F$ straightforwardly yields the excess of the mean logarithmic mass $\delta m$ that contributes to the surplus of the all-particle energy spectrum flux beyond the knee.}
          \label{tab:sup_iron}
          \begin{tabular}{cc} 
          \\
          \hline
          E~(PeV)   &   $m_{\Delta F}$      \\
	 \hline
	  11.2       &   $3.449 \pm 0.900$   \\      
	  14.1       &   $4.187 \pm 0.730$   \\
	  17.8       &   $3.608 \pm 0.394$   \\
	  22.4       &   $4.106 \pm 0.399$   \\
	  28.2       &   $3.886 \pm 0.332$   \\
	 \hline
         \end{tabular}
         \label{tab8}
\end{table}

\end{document}